\documentclass[twocolumn,showpacs,amssymb,aps,prc,nofootinbib,floatfix,superscriptaddress]{revtex4-1}
\usepackage{amsmath,graphicx,color,ulem}
\usepackage{hyperref}
\newcommand{\trento}{T$\mathrel{\protect\raisebox{-2.1pt}{R}}$ENTo}

\begin{document}
\title{Correlation between particle spectra and elliptic flow}

\author{Tribhuban Parida}
\affiliation{Department of Physical Sciences,
Indian Institute of Science Education and Research Berhampur, Transit Campus (Govt ITI), Berhampur-760010, Odisha, India}
\author{Rupam Samanta}
\affiliation{Institute of Nuclear Physics, Polish Academy of Sciences,  31-342 Cracow, Poland}
\author{Jean-Yves Ollitrault}
\affiliation{Universit\'e Paris Saclay, CNRS, CEA, Institut de physique th\'eorique, 91191 Gif-sur-Yvette, France}

\begin{abstract} 
  We introduce a new observable to probe the collective nature of the radial expansion of the quark-gluon plasma.
  This observable, dubbed $v_{02}(p_T)$, represents the correlation of the spectrum with elliptic flow, in the same way as the recently measured $v_0(p_T)$ represents the correlation of the spectrum with the transverse momentum per particle.
  The advantage of $v_{02}(p_T)$ over $v_0(p_T)$ is that it is measured using a three-particle cumulant, as opposed to a pair correlation, which significantly reduces the sensitivity to nonflow effects.
  We predict non-trivial differences between $v_{02}(p_T)$ and $v_0(p_T)$ in semi-central Pb+Pb collisions at the Large Hadron Collider (LHC) on the basis of hydrodynamic simulations. 
  A hint of these differences can be seen in the modification of $p_T$ spectra observed by ALICE in event-shape-engineered events. 
\end{abstract}

\maketitle

\section{Introduction}
The hadrons produced in an ultrarelativistic nucleus-nucleus collision are the remains of an expanding fluid, the quark-gluon plasma~\cite{Busza:2018rrf}, which is formed in the early stages of the collision. 
The fluid is in local thermal equilibrium, at least to a reasonable approximation, and cools down until it fragments into hadrons at a temperature $T_F\approx 160$~MeV~\cite{Andronic:2017pug}.
The momentum distribution of hadrons is thermal in the rest frame of the fluid~\cite{Cooper:1974mv}, therefore, spectra of outgoing hadrons are determined by the distribution of the fluid velocity, which is in turn determined by the energy density profile shortly after the collision (initial density). 

Hadron spectra fluctuate event by event, beyond the statistical fluctuations arising from the finite particle multiplicity. 
Our goal is to isolate these so-called ``dynamical'' fluctuations, and more specifically, those which are of collective origin. 
The first step in isolating collective effects is to look for correlations between particles separated by a rapidity gap~\cite{PHENIX:2003qra}, for the following reason.  
While the fluid velocity fluctuates event by event, its projection onto the transverse plane is independent of rapidity in every event, because the initial density is itself independent of rapidity~\cite{Dumitru:2008wn}. 
This approximate boost invariance~\cite{Bjorken:1982qr} is an essential symmetry of ultrarelativistic collisions. 
Along with collective flow, it gives rise to specific long-range correlations, in the sense that they span a wide rapidity interval~\cite{Nagle:2018nvi}. 

Therefore, in order to isolate the contribution of the fluid to the particle spectrum, one must correlate the spectrum with a property of the fluid measured at another rapidity. 
The simplest such measure is a two-particle correlation, where the spectrum is correlated with the transverse momentum per particle $[p_T]$. 
The corresponding observable, dubbed $v_0(p_T)$~\cite{Schenke:2020uqq}, has been recently measured by the ATLAS~\cite{ATLAS:2025ztg} and ALICE~\cite{ALICE:2025iud} collaborations in Pb+Pb collisions at the LHC. 
A well-known limitation of two-particle correlations is that they get sizable contributions from ``nonflow'' correlations~\cite{Borghini:2000cm,Feng:2024eos} which are not of collective origin. 
Implementing a rapidity gap largely reduces nonflow effects, but they are still present at high $p_T$, as a consequence of back-to-back jet correlations. 
It is likely that the centrality dependence of $v_0(p_T)$  observed by ATLAS at high $p_T$~\cite{ATLAS:2025ztg} is, at least in part, due to nonflow effects, as well as the differences between ALICE and ATLAS results. 

A systematic way of suppressing nonflow effects is to go beyond pair correlations and analyze multi-particle cumulants~\cite{Borghini:2000sa}. 
Their sensitivity to nonflow decreases by successive inverse powers of the event multiplicity as the order of the correlation increases. 
We propose to analyze the spectrum fluctuation using a three-particle cumulant. 
The idea is to correlate the spectrum with elliptic flow, $v_2$~\cite{STAR:2000ekf,ALICE:2010suc}, rather than with the momentum per particle. 
We denote the new observable as $v_{02}(p_T)$. 

We define $v_{02}(p_T)$ in Sec.~\ref{s:definitions}.
We show in Sec.~\ref{s:bozek} that it is a differential version of Bo\.zek's correlator between $[p_T]$ and $v_2$~\cite{Bozek:2016yoj}, which has been thoroughly studied at RHIC and LHC~\cite{ATLAS:2019pvn,ALICE:2021gxt,ATLAS:2022dov,STAR:2024wgy,CMS:2024rvk,STAR:2024ngl}. 
In Sec.~\ref{s:hydro}, we evaluate $v_{02}(p_T)$ in hydrodynamic simulations of mid-central Pb+Pb collisions at LHC energy and compare it with $v_0(p_T)$. 
The qualitative differences between $v_{02}(p_T)$ and $v_0(p_T)$ are interpreted in Sec.~\ref{s:twocomponent} within a two-component model, where the spectrum is determined by $[p_T]$ and $v_2$. 
In Sec.~\ref{s:ese}, we discuss how the dependence of spectra on event shapes observed by ALICE a decade ago~\cite{ALICE:2015lib} can be related to $v_{02}(p_T)$, and we compare experimental data with hydrodynamic simulations. 
 
\section{Definitions}
\label{s:definitions}

In this Section, we define $v_{02}(p_T)$ assuming that every collision event is a continuous fluid whose properties are independent of rapidity~\cite{Bjorken:1982qr}.
In such a description, the single-particle distribution of a given hadron species is a smooth function of momentum, and particles are independent in every event. 
The implementation of the method in an actual experiment is more complicated, due to the discretization (finite multiplicity) and to nonflow correlations.
These analysis details are relegated to Appendix~\ref{s:analysis}. 

Elliptic flow, $v_2$, is defined in every event as the modulus of the second Fourier coefficient of the azimuthal distribution of outgoing hadrons~\cite{Qiu:2011iv}.  
We denote by $n(p_T)$ the $p_T$ spectrum normalized on an event-by-event basis~\cite{ATLAS:2025ztg}, that is, 
\begin{equation}
\label{normalization}
\int_{p_T} n(p_T)=1,
\end{equation}
where $\int_{p_T}$ denotes the sum over $p_T$ bins, and $n(p_T)$ the fraction of particles in each bin.\footnote{In the case where the analysis is carried out for identified particles, one should sum over particles species {\it and\/} $p_T$ bins.  This rule applies throughout this paper, and a summation over particle species is always  implied when we integrate over $p_T$.}
The new observable is the relative correlation between $n(p_T)$ and $v_2^2$: 
\begin{equation}
  \label{defv02pt}
  v_{02}(p_T) \equiv \frac{\langle n(p_T)v_2^2\rangle-\langle n(p_T)\rangle\langle v_2^2\rangle}{\langle n(p_T)\rangle\langle v_2^2\rangle},
\end{equation}
where angular brackets denote the average over events in a centrality class.
This observable is analogous to $v_0(p_T)$~\cite{Schenke:2020uqq}, which is defined by 
\begin{equation}
  \label{defv0pt}
  v_{0}(p_T) \equiv \frac{\langle n(p_T)[p_T]\rangle-\langle n(p_T)\rangle\langle p_T\rangle}{\langle n(p_T)\rangle\sigma_{p_T}},
\end{equation}
where $[p_T]$ is the event mean $p_T$, and $\langle p_T\rangle$ and $\sigma_{p_T}$ are the expected value and standard deviation of $[p_T]$: 
\begin{align}
[p_T]&\equiv \int_{p_T} p_T n(p_T)\label{ventpt}\\
\langle p_T\rangle&\equiv \int_{p_T} p_T\langle n(p_T)\rangle\label{meanpt}\\
\sigma_{p_T}&\equiv \sqrt{\langle [p_T]^2\rangle-\langle p_T\rangle^2}. \label{sigmapt}
\end{align}
The normalization (\ref{normalization}) implies the sum rules
\begin{equation}
  \label{sumrule0}
  \int_{p_T} 
\langle n(p_T)\rangle  v_{02}(p_T)=
  \int_{p_T} 
\langle n(p_T)\rangle  v_{0}(p_T) =0. 
\end{equation}
By analogy with the definition of $v_0$~\cite{Schenke:2020uqq,Parida:2024ckk}, we define an integrated measure of $v_{02}(p_T)$, which we denote by $v_{02}$, by multiplying with $p_T/\langle p_T\rangle$ before integrating over $p_T$: 
\begin{align}
v_{02}&\equiv  \frac{1}{\langle p_T\rangle}\int_{p_T} p_T\langle n(p_T)\rangle v_{02}(p_T)  \label{defv02}\\
v_{0}&\equiv  \frac{1}{\langle p_T\rangle}\int_{p_T} p_T\langle n(p_T)\rangle v_{0}(p_T)=\frac{\sigma_{p_T}}{\langle p_T\rangle}.\label{defv0}
\end{align}
Inserting Eq.~(\ref{defv02pt}) into Eq.~(\ref{defv02}), one obtains
\begin{equation}
\label{v02meanpt}
v_{02}=\frac{\langle [p_T] v_2^2\rangle-\langle p_T\rangle\langle v_2^2\rangle}{\langle p_T\rangle\langle v_2^2\rangle}. 
\end{equation}

\section{Relation with Bo\.zek's correlator}
\label{s:bozek}
Eq.~(\ref{v02meanpt}) shows that $v_{02}$ is proportional to the correlation between $[p_T]$ and $v_2^2$. 
This correlation has been previously introduced by Bo\.zek~\cite{Bozek:2016yoj} in the form of the Pearson correlation coefficient:
\begin{equation}
\label{defrho2}
\rho_2\equiv\frac{ \langle [p_T] v_2^2\rangle - \langle p_T \rangle\langle v_2^2\rangle }
{\sigma_{p_T}\sigma_{v_2^2}}, 
\end{equation}
where $\sigma_{v_2^2}$ denotes the standard deviation of $v_2^2$: 
\begin{equation}
\sigma_{v_2^2}\equiv \left(\langle v_2^4\rangle -\langle v_2^2\rangle^2 \right)^{1/2}.
\label{defsigma}
\end{equation}
The numerators of Eq.~(\ref{v02meanpt}) and (\ref{defrho2}) are identical, while the denominators differ.
In Eq.~(\ref{v02meanpt}), we have chosen to normalize the correlation by the uncorrelated part (product of means), while in Eq.~(\ref{defrho2}) it is normalized by the product of standard deviations. 
Up to this different normalization, $v_{02}$ and $\rho_2$ are equivalent observables.  
Thus $v_{02}(p_T)$, which is the differential observable associated with $v_{02}$, can be viewed as a differential version of $\rho_2$.\footnote{We choose to make $\rho_2$ differential in the spectrum. It can also be made differential in elliptic flow~\cite{Samanta:2023rbn}.}

We now detail the relation between $v_{02}$ and $\rho_2$. 
Comparing Eq.~(\ref{defrho2}) with Eq.~(\ref{v02meanpt}), one obtains: 
\begin{equation}
\label{v02vsrho2}
v_{02}=v_0 \, \frac{\sigma_{v_2^2}}{\langle v_2^2\rangle}\,\rho_2,
\end{equation}
where $v_0$ is defined by Eq.~(\ref{defv0}), and the second term in Eq.~(\ref{v02vsrho2}) is the relative fluctuation of $v_2^2$, which is typically inferred from the fourth cumulant $v_2\{4\}$~\cite{Borghini:2001vi,Bozek:2016yoj}:
\begin{equation}
\label{sigv2rel}
\frac{\sigma_{v_2^2}}{\langle v_2^2\rangle}=\left(1-\frac{v_2\{4\}^4}{v_2\{2\}^4}\right)^{1/2}.
\end{equation}
This equation shows that in order to evaluate $\rho_2$, one must analyze $v_2\{4\}$, which is obtained from a cumulant of order four.
Therefore, $\rho_2$ is potentially more difficult to analyze than $v_{02}$, which only involves a cumulant of order three.

Let us evaluate the order of magnitude of $v_{02}$ in Pb+Pb collisions at the LHC, based on Eq.~(\ref{v02vsrho2}). 
In ultra-central collisions, both $\rho_2$~\cite{ATLAS:2019pvn} and $v_0$~\cite{ATLAS:2025ztg} decrease relative to mid-central collisions. 
In peripheral collisions, $\rho_2$ changes sign~\cite{ATLAS:2019pvn}. 
Therefore, $v_{02}$ is largest for mid-central collisions. 
We expect that the analysis of $v_{02}(p_T)$ will be easiest there, and we focus on the 30\%--40\% centrality window in the remainder of this paper. 
In this centrality range, using $\rho_2\approx 0.2$~\cite{ALICE:2021gxt}, $v_0\approx 0.02$~\cite{ATLAS:2025ztg},
$\sigma_{v_2^2}/\langle v_2^2\rangle\approx 0.7$~\cite{Giacalone:2017uqx}, we expect $v_{02}\approx 0.3\%$. 

\section{Predictions from hydrodynamic simulations}
\label{s:hydro}

We now evaluate $v_{02}(p_T)$ in hydrodynamic simulations.
We simulate 3000 Pb+Pb collisions at $\sqrt{s_{NN}}=5.02$~TeV in the 30\%--40\% centrality window.
We use the initial entropy as a centrality classifier. 
The initial entropy density is given by the \trento{} model of initial conditions~\cite{Moreland:2014oya}, without nucleon substructure and with a nucleon width $w=0.5$~fm. 
Within this model, we implement the usual prescription $p=0$, which corresponds to a density profile proportional to $\sqrt{T_AT_B}$, where $T_A$ and $T_B$ are the thickness functions of colliding nuclei. 
The fluctuation parameter is $k=1$. 
We normalize the density profile so as to reproduce the final multiplicity measured by ALICE in the 0\%--5\% centrality window~\cite{ALICE:2015juo}.

The hydrodynamic evolution starts at proper time $\tau_0=0.4$~fm$/c$. 
We solve hydrodynamic equations  using the publicly available MUSIC code~\cite{Schenke:2010nt,Schenke:2010rr,Paquet:2015lta}, with the Huovinen-Petreczky lattice-QCD equation of state with partial chemical equilibrium~\cite{Huovinen:2009yb}. 
We assume a minimal shear viscosity over entropy ratio $\eta/s=0.08$~\cite{Romatschke:2007mq} and zero bulk viscosity. 
We perform particlization following the Cooper-Frye prescription~\cite{Cooper:1974mv} on a freezeout hypersurface at $T=130$~MeV (corresponding to an energy density $0.18$~GeV/fm$^3$). 

Note that our hydrodynamic setup does not quite represent the state of the art: we neglect the initial transverse flow~\cite{Vredevoogd:2008id,Kurkela:2018wud} and hadronic rescatterings after freeze-out~\cite{Schenke:2020mbo}.
In addition, some of our model parameters, in particular the nucleon width~\cite{Giacalone:2022hnz} and the transport coefficients, are not tuned to the values favored by model to data comparisons~\cite{Bernhard:2019bmu,JETSCAPE:2020mzn,Nijs:2020roc}. 
Our goal is to illustrate some expected qualitative trends, rather than carry out systematic quantitative predictions, which will be done in a forthcoming presentation.
We expect that these qualitative trends are robust with respect to model details, as for the vast majority of flow observables. 

\begin{figure}[th!]
    \includegraphics[width=\linewidth]{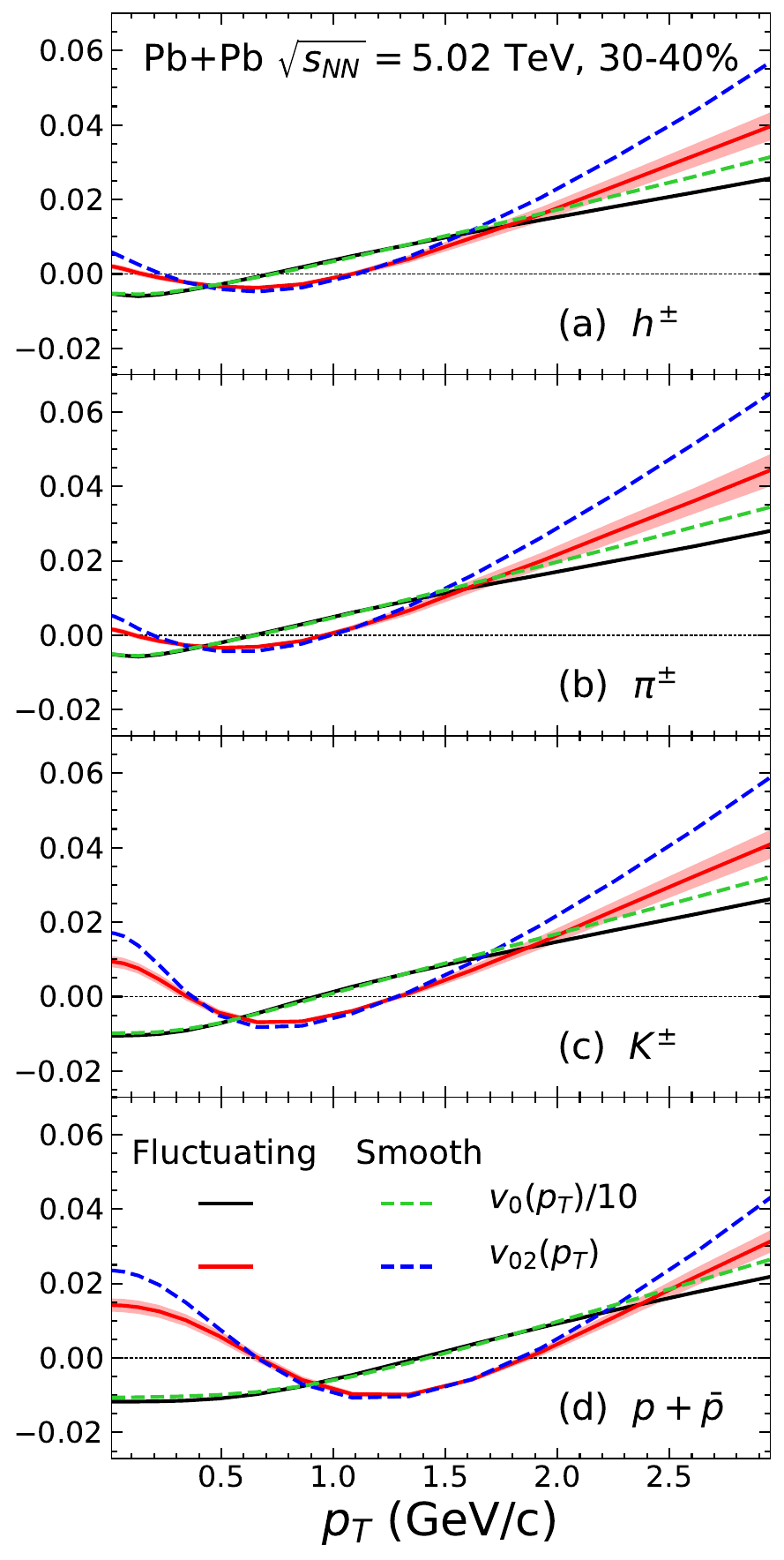}
    \caption{Hydrodynamic results for unidentified charged hadrons (a), pions (b), kaons (c) and protons (d).
      Solid red lines: $v_{02}(p_T)$, defined by Eq.~(\ref{defv02pt}), from our hydrodynamic simulation, where the band is the statistical error bar due to the finite number of events.  
Solid black lines: $v_0(p_T)$ defined by Eq.~(\ref{defv0pt}), scaled down by a factor $10$. 
Dashed lines are the result of the two-component model (Eqs.~(\ref{decv02}) and (\ref{decv0})), where the response functions $\alpha(p_T)$ and $\beta(p_T)$ are evaluated by solving hydrodynamics with smooth initial density profiles (see text). 
}      
        \label{fig:v02pthydro}
\end{figure}

The values of $v_{02}(p_T)$ returned by our hydrodynamic simulations are displayed in Fig.~\ref{fig:v02pthydro} for unidentified and identified charged hadrons.
In our simulations, ``unidentified charged hadrons'' stands for pions, kaons or protons. 
We also evaluate $v_0(p_T)$, which is displayed for the sake of comparison. 
One sees that the $p_T$ dependence differs for the two observables. 
While $v_0(p_T)$ changes sign once and is negative at small $p_T$, $v_{02}(p_T)$ changes sign twice and is positive at small $p_T$. 
The trend is more pronounced for heavier particles, which have larger $v_{02}(p_T)$ at small $p_T$.
Thus we predict an inversion of the standard mass ordering observed for elliptic flow $v_2(p_T)$~\cite{Huovinen:2001cy} and for $v_0(p_T)$~\cite{ALICE:2025iud}.

\section{Two-component model}
\label{s:twocomponent}

We now interpret these qualitative differences between $v_{02}(p_T)$ and $v_0(p_T)$ within a simple model.
We assume that the spectrum is solely determined by $[p_T]$ and $v_2$. 
Carrying out a fluctuation decomposition $n(p_T)=\langle n(p_T)\rangle+\delta n(p_T)$, $[p_T]=\langle p_T\rangle+\delta p_T$ and $v_2^2=\langle v_2^2\rangle+\delta v_2^2$ and linearizing in the fluctuations, one obtains\footnote{The reason why we consider the square of elliptic flow $v_2^2$ rather than $v_2$ itself is that $v_2$ is, strictly speaking, a complex Fourier coefficient~\cite{Qiu:2011iv}. For symmetry reasons, the modification of a scalar, such as the spectrum, can only appear at quadratic order.}
\begin{equation}
\label{decomposition}
\frac{\delta n(p_T)}{\langle n(p_T)\rangle}=\alpha(p_T)\frac{\delta p_T}{\langle p_T\rangle}+\beta(p_T)\delta v_2^2,
\end{equation}
where $\alpha(p_T)$ and $\beta(p_T)$ are response functions which are identical for all events in the centrality class.
The first term in the right-hand side is the modification of the spectrum induced by a small variation of $[p_T]$ at constant $v_2$, while the second term represents the change of the spectrum upon a variation of $v_2$ which leaves $[p_T]$ unchanged.

The spectrum fluctuation satisfies the sum rules $\int_{p_T}\delta n(p_T)=0$ and $\int_{p_T}p_T\delta n(p_T)=\delta p_T$. 
If Eq.~(\ref{decomposition}) holds for arbitrary (but small) values of $\delta p_T$ and $\delta v_2^2$, this in turn implies the sum rules: 
\begin{align}
    \int_{p_T}  \langle n(p_T)\rangle\alpha(p_T)= \int_{p_T}  \langle n(p_T)\rangle\beta(p_T)&=0\label{sumruleresp1}\\
      \int_{p_T} p_T \langle n(p_T)\rangle\alpha(p_T)&=\langle p_T\rangle\label{sumruleresp2}\\  
    \int_{p_T} p_T \langle n(p_T)\rangle\beta(p_T)&=0.\label{sumruleresp3}
\end{align}
Eq.~(\ref{sumruleresp1}) implies that $\alpha(p_T)$ changes sign at least once. 
If it changes sign only once, Eq.~(\ref{sumruleresp2}) implies that it is negative at small $p_T$ and positive at large $p_T$. 
Similarly, Eqs.~(\ref{sumruleresp1}) and (\ref{sumruleresp3}) imply that $\beta(p_T)$ changes sign at least twice.\footnote{Strictly speaking, these mathematical statements about changes of sign only hold for unidentified charged hadrons, not individually for each hadron species.}
This sign can easily be guessed: 
A larger elliptic flow goes along with steeper pressure gradients, which enhance particle production at large $p_T$.
One therefore expects $\beta(p_T)$ to be positive at large $p_T$. If it undergoes two successive changes of sign as $p_T$ decreases, this implies that it is positive again at small $p_T$. 

We now express $v_{02}(p_T)$ and $v_0(p_T)$ in terms of the response functions $\alpha(p_T)$ and $\beta(p_T)$. 
Multiplying Eq.~(\ref{decomposition}) by $\delta v_2^2/\langle v_2^2\rangle$, and averaging over events, we obtain:  
\begin{align}
v_{02}(p_T)&=\frac{\langle\delta n(p_T)\delta v_2^2\rangle}{\langle n(p_T)\rangle\langle v_2^2\rangle}\nonumber\\
&=\alpha(p_T)\frac{\langle\delta p_T\delta v_2^2\rangle}{\langle p_T\rangle\langle v_2^2\rangle}+\beta(p_T)\frac{\langle(\delta v_2^2)^2\rangle}{\langle v_2^2\rangle}\nonumber\\
&=\alpha(p_T)v_{02}+\beta(p_T)\frac{\langle v_2^4\rangle-\langle v_2^2\rangle^2}{\langle v_2^2\rangle}.
\label{decv02}
\end{align}
Similarly, multiplying Eq.~(\ref{decomposition}) by $\delta p_T/\sigma_{p_T}$ and averaging over events, we obtain:
\begin{align}
v_{0}(p_T)&=\frac{\langle\delta n(p_T)\delta p_T\rangle}{\langle n(p_T)\rangle\sigma_{p_T}}\nonumber\\
&=\alpha(p_T)\frac{\langle\delta p_T^2\rangle}{\langle p_T\rangle\sigma_{p_T}}
+\beta(p_T)\frac{\langle\delta v_2^2\delta p_T\rangle}{\sigma_{p_T}}\nonumber\\
&=\alpha(p_T)v_{0}+\beta(p_T)\rho_2\sigma_{v_2^2},
\label{decv0}
\end{align}
where we have used Eq.~(\ref{defrho2}). 
Thus we have expressed  $v_{02}(p_T)$ and $v_0(p_T)$ as linear combinations of the response functions $\alpha(p_T)$ and $\beta(p_T)$, but with different weights.
If $v_2$ and $[p_T]$ are uncorrelated, $v_{02}(p_T)$ is proportional to $\beta(p_T)$ while $v_0(p_T)$ is proportional to $\alpha(p_T)$:
This implies in particular that $v_0(p_T)$ is negative at small $p_T$ while $v_{02}(p_T)$ is positive, as observed in Fig.~\ref{fig:v02pthydro}. 
In practice, $v_2$ and $[p_T]$ are weakly correlated since $\rho_2$ is much smaller than unity. 
This simple two-component model thus naturally explains the qualitative differences between $v_{02}(p_T)$ and $v_0(p_T)$ seen in our simulations. 

We now carry out a quantitative calculation. 
In order to evaluate the response functions, we construct events with smooth initial density profiles, by averaging over a large number of fluctuating initial conditions~\cite{Song:2010mg}. 
With these smooth initial conditions, the only degrees of freedom are the overall normalization and the impact parameter. 
The fact that there are only two degrees of freedom implies that the two-component model strictly holds. 
Therefore, smooth events are the natural setup to evaluate response functions. 

We generate two smooth events with impact parameters $b=8.5$~fm and $b=10$~fm, corresponding roughly to the boundaries of the data shown in Fig.~\ref{fig:v02pthydro} (30\% and $40\%$). 
We adjust the normalization at $b=8.5$~fm so that the value of $[p_T]$ of outgoing hadrons matches the value of $\langle p_T\rangle$ for the full hydrodynamic simulation described above, namely, $696$~MeV/$c$. 
We use the same normalization constant for $b=10$~fm, resulting in a value of $[p_T]$ which is not exactly identical, but relatively close. 
We then rescale the initial  normalization at $b=8.5$~fm in such a way that $[p_T]$ is smaller by $\sim 10$~MeV/$c$. 
We thus obtain three different events, for which we evaluate $v_2$, $[p_T]$ and the spectra $n(p_T)$ after the hydrodynamic evolution. 
We eventually obtain $\alpha(p_T)$ and $\beta(p_T)$ by solving Eq.~(\ref{decomposition}) for the two variations (variation of normalization, and variation of impact parameter).

We finally inject these response functions into Eqs.~(\ref{decv02}) and (\ref{decv0}), and evaluate the $p_T$-integrated quantities in the right-hand side using the event-by-event simulation presented in Sec.~\ref{s:hydro}.  
The results are displayed as dashed lines in Fig.~\ref{fig:v02pthydro}. 
The two-component model quantitatively explains the differences between $v_{02}(p_T)$ and $v_0(p_T)$. 
The weights of $\alpha(p_T)$ and $\beta(p_T)$ are of the same order of magnitude in $v_{02}(p_T)$, whereas the weight of $\beta(p_T)$ in $v_0(p_T)$ is practically negligible. 
Therefore, the different $p_T$ dependence of $v_{02}(p_T)$ and $v_0(p_T)$ is mostly due to the response coefficient $\beta(p_T)$, as anticipated.  
The discrepancies between the two-component model and the full simulation can be attributed to the fact that we evaluate the response functions using smooth initial conditions.
The generic effects of fluctuating initial conditions on observables have long been known~\cite{Andrade:2008xh}.
They enhance particle production but reduce elliptic flow $v_2(p_T)$ at large $p_T$.
More generally, they reduce collective effects at large $p_T$.
The results displayed in Fig.~\ref{fig:v02pthydro} show that this reduction also applies to $v_0(p_T)$~\cite{Parida:2024ckk} and $v_{02}(p_T)$. 

\section{Relation with event-shape-engineering analyses}
\label{s:ese}

Event-shape engineering (ESE)~\cite{Schukraft:2012ah} is an intuitive way of studying the correlation of observables with elliptic flow.
The idea is to classify events, in a given centrality window, according to the value of a quantity $q_2$, defined from the directions of outgoing particles, which is on average larger for events with larger $v_2$.
One then studies how various observables depend on $q_2$. 
This technique has mostly been used in relation with the search for the chiral magnetic effect~\cite{Voloshin:2010ut,ALICE:2017sss,CMS:2017lrw}, but has been also applied to other observables, such as elliptic flow of charmed mesons~\cite{ALICE:2020iug} and jet quenching~\cite{ALICE:2023dei}.

\begin{figure}[th!]
    \includegraphics[width=\linewidth]{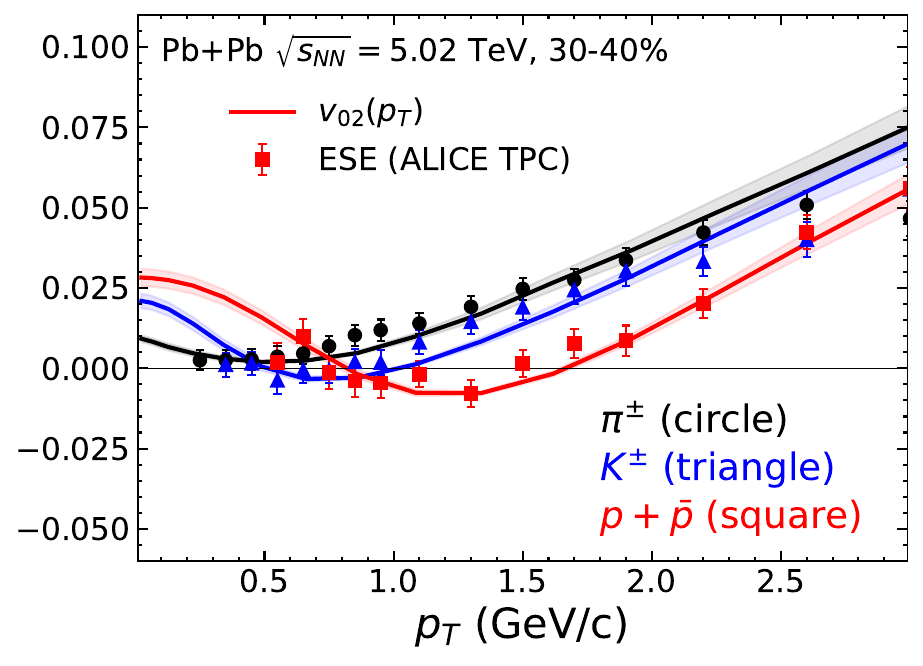}
    \caption{
      Symbols: relative difference between spectra of large-$q_2$ and small-$q_2$ events in Pb+Pb collisions at $\sqrt{s_{NN}}=2.76$~TeV~\cite{ALICE:2015lib} in the 30\%--40\% centrality window, where $q_2$ is the flow vector evaluated for particles seen in the Time Projection Chamber (TPC). 
      Lines: Fits using Eq.~(\ref{linearfit}), where $v_{02}(p_T)$ is taken from Fig.~\ref{fig:v02pthydro}.
        \label{fig:ESE}}
\end{figure}

The first application of the ESE technique by the ALICE Collaboration was to hadron spectra~\cite{ALICE:2015lib}. 
They measured their relative change from small-$q_2$ to large-$q_2$ Pb+Pb collision events.
This amounts to studying the correlation of spectra with elliptic flow, so that this ten-year-old analysis pioneered the study of $v_{02}(p_T)$. 
However, the ESE results cannot be directly compared with $v_{02}(p_T)$ as defined in Sec.~\ref{s:definitions} for two reasons: 
First, the dependence of a given observable on $q_2$ is proportional to the correlation of that observable with $v_2$, but the proportionality constant depends on the detector. 
Second, we define $v_{02}(p_T)$ using the {\it fraction\/} of particles $n(p_T)$ in the $p_T$ bin, as opposed to the {\it number\/} of particles $N(p_T)$.   
This amounts to normalizing the spectra on an event-by-event basis, and such a normalization is not implemented in the ESE analysis.
If one replaces $n(p_T)$ with $N(p_T)$ in the definition of $v_{02}(p_T)$, it is shifted by an additive constant proportional to the correlation between the total multiplicity and $v_2^2$.  
With these two differences taken into account, one expects:
\begin{equation}
  \frac{\frac{dN}{dp_Td\eta}({\rm large\ }q_2)-\frac{dN}{dp_Td\eta}({\rm small\ }q_2)}
       {\frac{dN}{dp_Td\eta}({\rm unbiased})}=av_{02}(p_T) +b,
       \label{linearfit}
\end{equation}
where $a$ and $b$ are dimensionless constants, which are independent of $p_T$ and particle species, but depend on analysis details.

Fig.~\ref{fig:ESE} displays our fit to ALICE data using Eq.~(\ref{linearfit}), where $v_{02}(p_T)$ is taken from the hydrodynamic calculation in Fig.~\ref{fig:v02pthydro}.
We obtain a reasonable fit with $a=1.5$ and $b=0.007$. 
This comparison between theory and data shows that the hydrodynamic calculation simultaneously captures the $p_T$ dependences and the relative ordering of pion, kaon, proton results.
In particular, data seem to support the inverted mass ordering predicted by hydrodynamics at low $p_T$. 
The parameter $a$ is of order unity, which means that hydrodynamics naturally captures the order of magnitude of the observed effect. 

\section{Conclusions}
\label{s:discussion}

We have introduced a new observable $v_{02}(p_T)$, which probes collective flow through the correlation between hadron spectra and elliptic flow in heavy-ion collisions.
It is straightforward to measure, and the results can be directly compared to model calculations.
In this respect, it represents a significant improvement over the use of ESE events pioneered by ALICE~\cite{ALICE:2015lib}. 
$v_{02}(p_T)$ is a non-trivial probe of collectivity, in the sense that it would vanish in the absence of final-state interactions, in the same way as elliptic flow itself.
It is a three-particle cumulant and is therefore much less sensitive to nonflow effects than observables based on pair correlations, such as $v_0(p_T)$.
This will be useful in particular at high $p_T$, where there are large residual nonflow effects from jets, even after applying a rapidity gap. 

We predict that the $p_T$ dependence of $v_{02}(p_T)$ differs significantly from that of $v_0(p_T)$ in mid-central Pb+Pb collisions. 
In particular, while $v_0(p_T)$ for identified hadrons satisfies the usual mass ordering, the order is {\it inverted\/} for $v_{02}(p_T)$ at small $p_T$.
This inversion results from a modification of the spectra which is induced specifically by a change of elliptic flow at fixed $[p_T]$. 
One typically expects that in central collisions, where elliptic flow is smaller, the usual mass ordering will be restored. 
Detailed hydrodynamic calculations are needed in order to predict at which centrality this effect appears, and whether it is also expected for smaller colliding nuclei. 

We thus expect that the $p_T$ dependence of $v_{02}(p_T)$ will change qualitatively as a function of centrality and system size.
This is a new feature, which is not observed with usual differential observables:
Particle spectra~\cite{Muncinelli:2024izj}, $v_0(p_T)$~\cite{ATLAS:2025ztg} and $v_n(p_T)$~\cite{ATLAS:2018ezv} are almost centrality independent, once properly rescaled. 
In hydrodynamics, they depend weakly on centrality {\it and\/} on transport coefficients after rescaling~\cite{Muncinelli:2024izj,Parida:2024ckk}. 
$v_{02}(p_T)$ is different because it results from the interplay between two different phenomena. 
We have modeled it as a linear combination of two different response functions, whose relative weight depends on centrality. 
Detailed hydrodynamic calculations are also needed to assess the dependence of each component on transport coefficients and initial conditions.

$v_{02}(p_T)$ thus reveals new phenomena, which probe the pressure gradients in the fluid in a more detailed way than existing observables. 
It can be measured at the LHC and at RHIC for various colliding systems. 
In the same way as the  $\rho_2$ correlator has been proven a sensitive probe of nuclear deformation~\cite{Giacalone:2020awm,Jia:2021wbq,Bally:2021qys,Jia:2021qyu,Bally:2023dxi,Zhao:2024lpc,Fortier:2024yxs,Mantysaari:2024uwn}, $\alpha$ clusters in light nuclei~\cite{Zhang:2024vkh} and nucleon width~\cite{Giacalone:2021clp}, we expect that its differential version $v_{02}(p_T)$ will bring new, tight constraints on nuclear structure. 
Its generalization to other Fourier harmonics, e.g. the correlation between spectra and triangular flow $v_{03}(p_T)$~\cite{Alver:2010gr}, is straightforward. 
We anticipate a rich upcoming activity on this topic. 

\begin{acknowledgments}
We thank J\"urgen Schukraft for discussions which inspired this work. 
RS is supported by the Polish National Science Centre grant: 2023/51/B/ST2/01625. 
\end{acknowledgments}

\appendix
\section{Experimental analysis} 
\label{s:analysis}

In this Appendix, we present specific suggestions on how to analyze $v_{02}(p_T)$ and $v_{02}$ defined by Eqs.~(\ref{defv02pt}) and (\ref{v02meanpt}) in heavy-ion experiments. 
We start with the denominators, which both involve $\langle v_2^2\rangle$.
This is a standard observable, which is inferred from a pair correlation:
\begin{equation}
  \label{defpair}
\langle v_2^2\rangle\equiv\langle XY\rangle_c\equiv\langle XY\rangle-\langle X\rangle\langle Y\rangle,
\end{equation}
where $X$ and $Y$ are two quantities which are measured in each event and angular brackets denote an average over events in a centrality class. 
The subscript $c$ denotes the connected part of the correlation, and the second term in the right-hand side subtracts out the uncorrelated (disconnected) part. 
 $X$ and $Y$ are defined in each event by 
\begin{eqnarray}
  \label{xyvnint}
X&\equiv&\frac{1}{N_A}\sum_{j=1}^{N_A}\exp(2i\varphi_j)\cr
Y&\equiv&\frac{1}{N_B}\sum_{j=1}^{N_B}\exp(-2i\varphi_j), 
\end{eqnarray}
where $N_A$ and $N_B$ denote the number of particles in two detectors (or two parts of the same detector) $A$ and $B$, and $\varphi_j$ their azimuthal angles.
In order to suppress nonflow correlations, most of which are short-range in rapidity~\cite{Dinh:1999mn,Borghini:2000cm}, $A$ and $B$ are usually separated by a gap in pseudorapidity~\cite{PHENIX:2003qra}.
One implicitly assumes that the underlying $v_2$ is the same in $A$ and $B$, or, equivalently, that longitudinal decorrelation~\cite{CMS:2015xmx} is a negligible effect~\cite{ATLAS:2017rij}. 

One usually carries out corrections in order to make the efficiencies in $A$ and $B$ perfectly isotropic~\cite{Poskanzer:1998yz}, so that the disconnected term  $\langle X\rangle\langle Y\rangle$ in Eq.~(\ref{defpair}) vanishes. 
Note, however, that the analysis works just as well if the detector does not have uniform acceptance~\cite{Borghini:2000sa}, provided that one properly subtracts the disconnected term.

We now explain how to measure the numerators of Eqs.~(\ref{defv02pt}) and (\ref{v02meanpt}).
They are three-particle cumulants of the form 
\begin{widetext}
\begin{equation}
  \label{def32}
  \langle X Y Z\rangle_c= \langle XYZ\rangle
  -\langle XY\rangle\langle Z\rangle-\langle YZ\rangle\langle X\rangle-\langle ZX\rangle\langle Y\rangle+2\langle X\rangle\langle Y\rangle\langle Z\rangle, 
\end{equation}
\end{widetext}
where $X$, $Y$ are defined by Eq.~(\ref{xyvnint}), and $Z$ is a third quantity measured in a third detector $C$.
(Whether $C$ should be separate from $A$ and $B$ will be discussed in detail below.)

For $v_{02}(p_T)$, $Z$ is the fraction of particles seen in $C$ which belong to the $p_T$ bin: 
\begin{equation}
\label{zpt}
Z(p_T)\equiv\frac{N_C(p_T)}{N_C}. 
\end{equation}
For $v_{02}$, $Z$ is the transverse momentum per particle in $C$:
\begin{equation}
\label{zint}
Z\equiv\frac{1}{N_C}\sum_{j=1}^{N_C}p_{T,j}. 
\end{equation}
When analyzing $v_{02}(p_T)$ for identified hadrons, the only change is that the numerator of Eq.~(\ref{zpt}) is the number of these identified hadrons. 
All other quantities ($X$, $Y$ and $N_C$ in the denominator of Eq.~(\ref{zpt})) should still be evaluated for unidentified charged hadrons. 
Note that if the detector has uniform acceptance and efficiency, azimuthal symmetry implies $\langle X\rangle=\langle Y\rangle=\langle XZ\rangle=\langle YZ\rangle=0$, so that Eq.~(\ref{def32}) simplifies to 
$\langle X Y Z\rangle_c= \langle XYZ\rangle-\langle XY\rangle\langle Z\rangle$.
If the acceptance is not uniform, one should keep all the terms in Eq.~(\ref{def32}). 

Consistency requires that $\langle n(p_T)\rangle$ and $\langle p_T\rangle$ in the denominators of Eqs.~(\ref{defv02pt}) and (\ref{v02meanpt}) are also measured in detector $C$.
Eventually, Eq.~(\ref{defv02pt}) is implemented as 
\begin{equation}
v_{02}(p_T)=\frac{\langle XYZ(p_T)\rangle_c}{\langle XY\rangle_c\langle Z(p_T)\rangle}. 
\end{equation}
$C$ is the detector where $v_{02}(p_T)$ is analyzed, while $A$ and $B$ are used as reference detectors.

\begin{figure}[th!]
    \includegraphics[width=\linewidth]{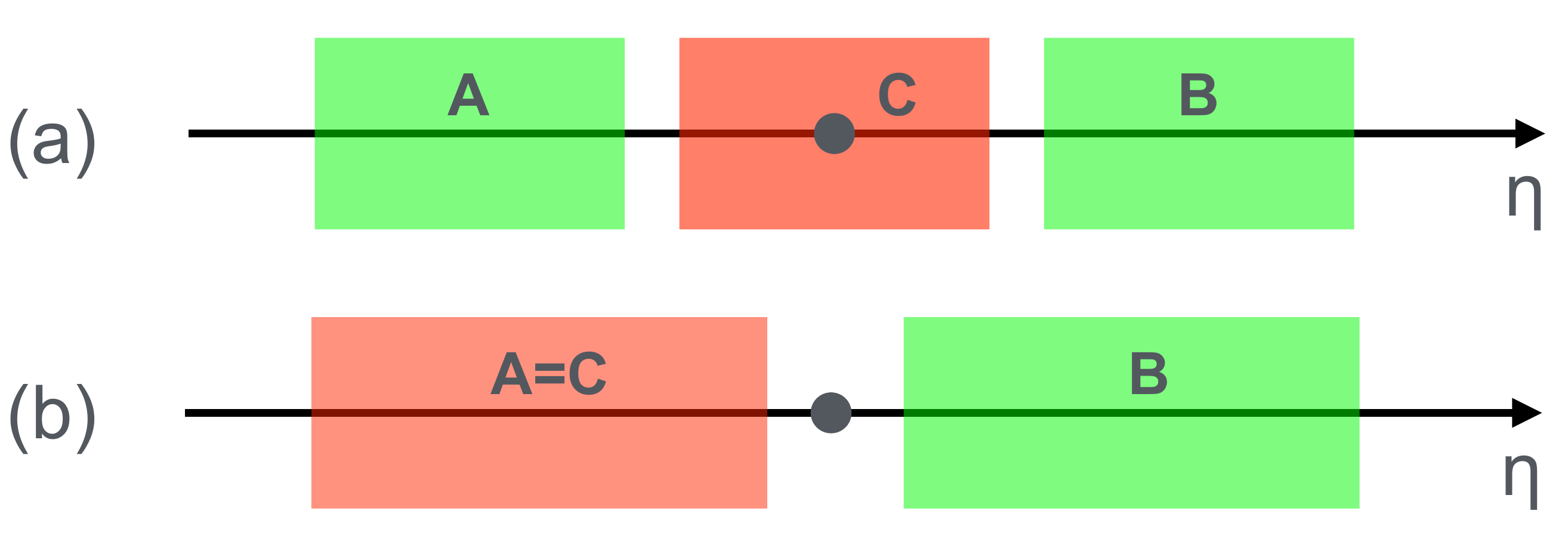}
    \caption{Schematic representation of two possible choices of subevents $A$, $B$ and $C$, defined by selections in pseudorapidity $\eta$.
            (a) Usual choice for $\rho_2$~\cite{ATLAS:2019pvn,ALICE:2021gxt,CMS:2024rvk}.
    (b) Choice consistent with the analysis of $v_0(p_T)$~\cite{ATLAS:2025ztg,ALICE:2025iud}.} 
        \label{fig:subevents}
\end{figure}

Let us finally comment on the choice of $C$.
Avoiding any overlap with $A$ and $B$ seems an optimal choice so as to reduce short-range nonflow correlations.
The standard choice for the analysis of $\rho_2$ is to have $A$ and $B$ symmetrically located around mid-rapidity, and $C$ around mid-rapidity, 
in the gap between $A$ and $B$~\cite{ATLAS:2019pvn,ALICE:2021gxt,CMS:2024rvk} (Fig.~\ref{fig:subevents} (a)). 

Since $v_{02}(p_T)$ is a differential version of $\rho_2$, it would seem natural to carry out the analysis in the exact same way. 
But it seems also important to have an apples-to-apples comparison between $v_{02}(p_T)$ to $v_0(p_T)$, in order to reveal the differences predicted by hydrodynamics, which are illustrated by Fig.~\ref{fig:v02pthydro}.  
Now, $v_0(p_T)$ has been obtained by correlating two subevents symmetrically located around mid-rapidity~\cite{ATLAS:2025ztg,ALICE:2025iud}, which amounts to choosing $C$ on one side of $\eta=0$, as opposed to around $\eta=0$.
In such a configuration, $C$ typically overlaps with $A$ or $B$ 
(Fig.~\ref{fig:subevents} (b)).
This is not a problem {\it per se\/}, because the sensitivity to nonflow effects is much reduced for a three-particle cumulant~\cite{Borghini:2000sa}, relative to a pair correlation. 
Note that the STAR analysis~\cite{STAR:2024wgy} of $\rho_2$ is carried out with {\it full overlap\/} $A=B=C$. 
They evaluate systematic errors from nonflow effects by repeating the analysis with separate $A$ and $B$, and $C=A\bigcup B$, and the difference is found to be small.

In the case of full or partial overlap between $A$, $B$ and $C$, one must remove self-correlations, which can be done systematically~\cite{DiFrancesco:2016srj}.
As an illustration, we write down this subtraction explicitly in the specific case where $A=C$ (Fig.~\ref{fig:subevents} (b)).
The particles appearing in $Z$, Eq.~(\ref{zpt}), are also present in $X$, Eq.~(\ref{xyvnint}). Self correlations correspond to the diagonal terms in the product $XZ$, which involves a double sum. 
We define in each event 
\begin{equation}
  \label{defq2}
  Q_2(p_T)\equiv \sum_{j=1}^{N_A(p_T)}\exp(2i\varphi_k),
\end{equation}
where the sum runs over all the particles belonging to the $p_T$ bin.
The subtraction of self-correlations amounts to replacing:
\begin{align}
  XZ&=\frac{N_A(p_T)\sum_{j=1}^{N_A}\exp(2i\varphi_j)}{N_A^2}\nonumber\\
  &\rightarrow
  \frac{N_A(p_T)\sum_{j=1}^{N_A}\exp(2i\varphi_j)-Q_2(p_T)}{N_A^2-N_A(p_T)}.
  \label{subtractself}
\end{align}
This change must be implemented in the right-hand side of Eq.~(\ref{def32}) in the terms $\langle XYZ\rangle$ and $\langle XZ\rangle$. 
Note that self-correlations are subtracted both in the numerator and in the denominator of Eq.~(\ref{subtractself})~\cite{STAR:2024wgy}.

\end{document}